\input aa.cmm
\input psfig.tex

\voffset=1 truecm
\overfullrule=0 pt

\MAINTITLE{The morphological segregation of galaxies in 
clusters.} 
\SUBTITLE{III. The distant cluster Cl0939+4713 (Abell 851)\FOOTNOTE{Based 
on observations made with the NASA/ESA {\it Hubble Space
Telescope}. Table 1 is only available in electronic form at the
CDS anonymous ftp to cdsarc.u-strasbg.fr (130.79.128.5) or via
http://cdsweb.u-strasbg.fr/Abstract.html}}

\AUTHOR{S. Andreon@1@2\FOOTNOTE{Present adress: Osservatorio di
Capodimonte, via Moiariello 16, 80131 Naples, Italy, e-mail: 
andreon at cerere.na.astro.it}, E. Davoust@2 and T. Heim@2@3}

\INSTITUTE{
@1 IFCTR-CNR, via Bassini 15, 20133 Milano, Italy \newline
\noindent 
@2 UMR 5572, Observatoire Midi-Pyr\'en\'ees, 14 Avenue E. Belin, 31400 
Toulouse, France, E-mail: davoust at obs-mip.fr\newline
\noindent 
@3 Ecole Nationale Sup\'erieure de Lyon, 46 All\'ees d'Italie, 69364 Lyon 
Cedex 07, France, E-mail: Thomas.Heim  at ens.ens-lyon.fr\newline}

\DATE{Received ..., accepted on November 26, 1996}

\ABSTRACT{
We have performed an isophotal analysis of galaxies in the distant
($z=0.4$) cluster 
Cl0939+4713 (Abell 851), using post-refurbished {\it Hubble Space Telescope} 
images.  Morphological type estimates for the galaxies are given. A rigorous 
comparison of the properties of the types shows that early-type galaxies in 
Cl0939+4713 are, within the present statistical or systematic errors, 
indistinguishable from their counterparts in Coma in all their studied 
properties, namely ellipticity profile, slope of the color-magnitude relation 
from the near-ultraviolet to the near-infrared colors, mean surface brightness 
and luminosity function in the restframe photographic $J$ band, homogeneity in 
color (around the color-magnitude relation). Furthermore ellipticals and 
lenticulars are separately homogeneous in their mean surface brightness, and 
have similar relative mean surface brightnesses in both clusters. 
 
Spirals are overabundant in Cl0939+4713 with respect to Coma, but twice
less than previously estimated, and are more similar to field spirals than
to cluster spirals. We suggest that the differences in the photometric
properties of the spirals in the two clusters arise from differences in 
cluster gas density distribution which ultimately bring star formation in the 
spirals to a stop in Coma, but not in Cl0939+4713. 

The morphological types are segregated along a privileged direction that 
coincides with the position angle of the major axis of the outer X-ray 
isophotes in the cluster, just like in Coma and Perseus. 
}

\THESAURUS{03((11.03.4 Cl0939+4713 (Abell 851);  
11.05.1; 11.06.2; 11.05.2; 11.12.2; 11.19.2)} 
\maketitle 
\titlea {Introduction}

Many observational efforts over a long period of time have been devoted to 
measuring the evolution of the properties of the Hubble types. This has mainly 
been done by comparing the properties of the types at large distances 
(z$\sim0.2-0.4$) with a nearby reference sample (see as exceptions, e.g. 
Courteau, de Jong \& Broeils 1996, Zepf 1996). 

When comparing the properties of the Hubble types, a first problem came from 
the difficulty, and in part the subjectivity, in estimating Hubble types. 
Expert morphologists are able to classify galaxies in morphological types 
with a reproducibility in the range 50 to 80 \% (Andreon \& Davoust 1996)
when galaxies are classified in three types.  The 
reproducibility is obviously lower when each morphologist uses his personal 
definitions for the types. Unfortunately, many recent studies of distant 
galaxies (e.g. Casertano et al.  1995, Glazebrook et al. 1995, Driver, 
Windhorst \& Griffits 1995), but not all (Dressler et al. 1994a,b, Abraham et 
al. 1996, van den Bergh et al. 1996) use definitions for the types which are 
not identical to those used by morphologists for nearby galaxies. 
Morphologists classify (nearby) galaxies by their resemblance to standards 
(Hubble 1936, de Vaucouleurs 1959, Sandage 1961), whereas distant galaxies are 
often classified without any reference to standards and in classes that are 
not easy to link to the traditional ones (for example Glazebrook et al. 1995 
classify usual edge-on lenticulars as spirals, and put together in one class 
compact galaxies and ellipticals). 

Second, the comparison of types at different redshifts has been hampered 
for a long time by the lack of angular resolution of groundbased telescopes 
that do not allow imaging distant galaxies with the same restframe spatial 
resolution as nearby ones, and therefore do not allow one to classify distant 
galaxies, in spite of great efforts to observe distant galaxies in 
sub-arcsecond seeing conditions (e.g. Thompson 1986, Lavery, Pierce \& McClure 
1992). The {\it Hubble Space Telescope} ({\it HST}) has now opened a new era 
and the situation has been partly reversed, since images of distant galaxies 
are now available with better restframe resolution than for nearby galaxies 
(see Sect. 3.1 for details). 

The goal of this paper is to study the galaxies in Cl0939+4713, a distant 
cluster observed by the {\it HST}, to compare their properties to the ones of 
galaxies in the Coma and Perseus clusters, presented in previous papers of 
this series (Andreon 1994, 1996a), in order to gain insight into the epoch 
dependence of the properties of the Hubble types. This is done using the same 
method for classifying all the galaxies from images of similar restframe 
resolution and for samples selected in the same restframe passband 
down to the same absolute magnitude.  

\medskip 

The paper is organized as follows. Section 2 outlines our present understanding 
of the properties of the galaxy populations in Cl0939+4713. Section 3 deals 
with the problems that could bias the comparison of morphological types 
of galaxies
at different redshifts. The morphological types of galaxies in Cl0939+4713 are 
presented and discussed in Section 4. In Section 5 we study the properties of 
the types, among themselves and with respect to local counterparts, and we 
devote Section 6 to a discussion and a summary of the results.  We adopt 
$H_0=50$ km s$^{-1}$ and $q_0=0.5$, but, when appropriate, we discuss the 
consequences of the adopted cosmological model. 

\titlea{The distant cluster Cl0939+4713}

Cl0939+4713 (Abell 851) is a distant cluster (at $z=0.407$) for which a large 
effort has been made to collect large samples of data at different 
wavelengths. Dressler \& Gunn (1992) give $g, r, i$ magnitudes, $g-r, r-i$ 
colors and positions. They also give the redshift of some galaxies. Dressler 
et al. (1994a, hereafter DOBG) give the morphological types of galaxies in the 
cluster core, as estimated from pre-refurbished {\it HST} images, and a first 
analysis of the properties of the Hubble types, revealing the ordinariness of 
these galaxies from the point of view of their morphology, from their location 
in the color-magnitude diagram, from the usual luminosity function and from 
the spatial distribution of the morphological types. 

Dressler et al. (1994b) present post-refurbished {\it HST} images of this 
cluster and re-classify the galaxies, but they do not give their morphological 
estimates. The comparison of the {\it HST}
morphological type estimates, before and after 
refurbishing, performed by Dressler et al. (1994b), shows that the two 
estimates differ for only 20 \% of the sample. 

In a narrow-band study, Belloni et al. (1995) and Belloni \& R\"oser (1996) 
point out the presence of numerous E+A galaxies, and that their radial 
distribution is less concentrated than those of early-type galaxies. 

Stanford, Eisenhardt \& Dickinson (1995) give the infrared colors of 
galaxies in an area that is much larger than the {\it HST} field of view, but 
limited to a bright magnitude ($K=18$ mag). They find that the redder 
galaxies are bluer than present day Es and S0s by a quantity which is 
compatible with the blueing expected for passive galaxy evolution. 
Furthermore,
the subsample of early-type galaxies with known morphological types has a 
small scatter around the color-magnitude relation in the visual and
near-infrared 
colors; this is indicative of the homogeneity of its photometric properties 
at these wavelengths. However, the early-type galaxies seem not to have the 
expected $J-K$ and $H-K$ colors, as they have brighter $H$ and lower $J$
fluxes for their $K$ magnitude.

In the effective radius vs effective surface brightness plane, the 14 galaxies 
of this cluster with known surface brightness parameters occupy the same locus 
as local early-type galaxies (Pahre, Djorgovski \& de Carvalho 1996).
Barrientos, Shade \& L\'op\`ez-Cruz (1996) instead suggest 
that, for a given size, early-type 
galaxies in Cl0939+4713 are brighter than the Coma ones, and 
note that this small discrepancy with the results of Pahre, Djorgovski \& de 
Carvalho (1996) is within the systematic errors of the two works. 

In the X-rays, the cluster looks irregular, as it is azimuthally asymmetric
(Schindler \& Wambsganss 1996). Seitz et al. (1996) present the 
two-dimensional mass distribution constructed from weak lensing on background
galaxies and find that mass follows light.

\titlea{Problems in the comparison of types at different redshifts}

Comparing the properties of a given Hubble type in different environments
is a non trivial task when very 
different observational materials are used for classifying the galaxies, and 
the images taken with the Wide Field Camera 2 of the {\it HST} differ in many 
respects from images taken with groundbased telescopes. The selection criteria 
and the classification methods may also differ. In the next sub-sections we 
detail the problems that arise because of these differences, and explain how 
we solved them. 

\titleb{The comparison sample}

The Hubble type estimates depend on the resolution of the images used for
classifying the galaxies (de Vaucouleurs \& Buta 1980, Poulain, Nieto \&
Davoust 1992). The resolution of the {\it HST} images for the Wide Field
Camera 2 is 0.1 arcsec (FWHM) which corresponds to 0.65 Kpc in  
the restframe of galaxies at $z\sim0.4$.  For a meaningful
comparison, such a
resolution in Kpc is needed for the nearby comparison samples as well.  

This perfect sample does not exist, so we settled for the next best 
thing,
two samples of galaxies in Coma (Andreon et al. 1996a,b). The two samples are 
composed of 187 galaxies brighter than  $M_J=- 19.8$ mag within one degree 
from the cluster center (hereafter the Coma-whole sample) and a deeper sample 
of 96 galaxies brighter than $M_J=-19.3$ mag within a central circular region 
of 15 arcmin in radius offcentered from the cluster center by 21 arcmin 
(hereafter the Coma-center sample).  These two samples exceed in size that of 
Saglia, Bender \& Dressler (1993). 

The angular resolution of the galaxy images is not uniform; it is broadly 
distributed, and centered at 0.78 Kpc. By comparing Hubble type estimates 
determined from images with different resolutions, some better than 0.65 Kpc, 
we have shown that the Hubble type estimate is not sensitive to small 
differences in resolution around 0.65 Kpc (Andreon et al. 1996b, Andreon \& 
Davoust 1996). The resolution of our comparison sample can thus be considered 
sufficient for our purpose. 

\titleb{Sampling and shape of the point spread function}

Sampling and contrast are two basic ingredients for recognizing
morphological components of galaxies, such as bar, disk, spiral structure
and, more generally, deviations of the isophotes from the perfect
elliptical shape. 

Even in the ideal case when the resolution in Kpc is the same for the nearby 
and the distant galaxies, the image qualities still differ, because the 
sampling and the shape of the point spread functions are not equal.  The FWHM 
of the {\it HST} point spread function is sampled with only one pixel, whereas 
the FWHM of the groundbased point spread function is sampled with typically 3 
to 4 pixels. This fact induces some differences in the Hubble type estimates 
for galaxies of the same Hubble type observed from the ground or from space. 

We have performed simple (i.e. noiseless) two-dimensional simulations of the 
appearance of structural components such as halo, bar, bulge, disk, etc. when 
observed with the Wide Field Camera 2 and from the ground (Andreon 1996b).  
They show that we do not lose {\it large} structural components, such as the 
disk, the envelope or the spiral pattern of distant galaxies (brighter than 
$M_J=-19.0$ mag) observed with the Wide Field Camera 2 more than we do for 
nearby galaxies observed from the ground. Therefore, galaxies of the same 
Hubble type will be put in the same coarse class (i.e. E, S0, S). 

But they also show that {\it small} details, such as a bar, or small 
deviations from the perfect elliptical shape, those allowing the 
classification of ellipticals as boxy or disky and of lenticulars as barred or 
unbarred, are more likely to be lost with {\it HST} images of distant galaxies 
than with groundbased images of nearby galaxies. Therefore, the relative 
proportions of ellipticals or lenticulars of the two subtypes measured in 
distant clusters are not comparable with those in nearby clusters. 

To be more specific, if the Coma barred lenticulars were artificially placed 
at the distance of Cl0939+4713, the major axis of their bar would be 0.3 to 
0.4 arcsec (i.e. 3 to 4 pixels) long on average.  It is very difficult to 
measure the change in position angle of such a short major axis and the 
simultaneous increase of ellipticity, which are distinctive features of a bar.  
Such result cannot be extrapolated to bars in spirals, which are 
morphologically different, and therefore to the bulk of the galaxies of the 
Hubble Deep Field. Incidentally, we note that only a very small fraction of 
barred galaxies has been detected in Cl0939+4713 by us (see column 10 of Table 
1). 

The arms of flocculent spirals are more likely to be lost in Wide Field 
Camera 2 images, as already noted by Ellis (1996), thus introducing 
another 
possible bias between groundbased and space observations.  But we have 
two reasons to think that the fraction of misclassified flocculent spirals is 
small: first of all, these galaxies often have HII regions, in which case they 
are correctly classified as spirals; second, flocculent spirals are not 
frequent in the center of clusters, even less so than grand design spirals (at 
least in clusters at zero redshift). 

\titleb{Pass-band: selection and absolute magnitude limit}

The Hubble type composition depends on the pass-band in which the sample has 
been selected, since samples selected in the blue are richer in spirals than 
those selected in the red. 

The galaxies in Cl0939+4713 were observed and selected through the 
$F702W$ filter, large (1950 \AA) and centered at $\sim$ 6900 \AA, (Holtzmann et 
al. 1995) which samples the restframe emission at $\lambda \sim$ 4900 \AA \ 
with a bandwidth of 1390 \AA. 

Galaxies in nearby clusters are often selected in the photographic $J$ filter 
(and in particular our major comparison sample, the Coma cluster, was selected 
in a $J$-like band), which is centered at 4600 \AA\ with a bandwidth of 1500 
\AA, approximately matching the emission sampled by the $F702W$ filter for 
distant galaxies.  Therefore we can use different filters for selecting 
samples in approximately the same restframe pass-band, as has been done, for 
example, by Arag\'on-Salamanca et al. (1993) and Rakos \& Schombert (1995) for 
other filters. 

The $F702W$ magnitudes for the galaxies in Cl0939+4713 were taken from
Seitz et al.'s (1996) catalogue (kindly made available by J.-P. Kneib). 
Two bright galaxies are clearly missing from Seitz et al.'s catalogue (DG
371, and DG 438; the names are from Dressler \& Gunn 1992).  These two
galaxies have been added to the catalogue and their $F702W$ magnitudes
have been estimated from the Gunn $r$ magnitudes assuming for their
$r-F702W$ color the median value (0.505 mag) of all the galaxies in
common between Seitz et al. (1996) and Dressler \& Gunn (1992). 

The $b_{GMP}$ magnitudes (which are magnitudes measured through a 
$J-$like filter but with the Johnson B photometric zero point)  
of the Coma galaxies listed by Godwin,
Metcalfe \& Peach (1983) are re-transformed to $J$ by the relation

$J=b_{GMP} - 0.57$ 

\noindent
computed from the comparison of the $b_{GMP}$ and $J$ magnitudes, the latter 
taken from Butcher \& Oemler (1984) for 148 galaxies in common with our sample 
of 187 with $b_{GMP}<16.5$. This transformation takes into account the 
difference in the zero point between the two systems, and it is not a color 
transformation, since both magnitudes map the same wavelength range. 

We now determine the zero-point shift between the $F702W_{z=0.4}$ and
$J_{z=0}$ systems, in order to select the distant and nearby samples down
to the same limiting absolute magnitude. 

We compute the $F702W-J$ color at redshift zero and the K corrections in 
$F702W$ by convolving the galaxy spectral energy distributions listed in 
Coleman, Wu \& Weedman (1980) with the $F702W$ (at 
http://www.stsci.edu/ftp/instrument) and $J_k$ (J.-M. Miralles, private 
communication) filter shapes, and by normalizing the $F702W-J$ color at 
redshift zero of ellipticals to the one listed in Fukugita, Shimasaku \& 
Ichikawa (1995). We find: 

$F702W_{z=0.4}-J_{z=0}=-0.95$ \quad $\sigma=0.1$ 

\noindent
where $\sigma$ indicates the scatter between the four different 
spectrophotometric types (E/S0, Sbc, Scd, Irr). Since there is a small 
residual dependence between the spectrophotometric type and the photometric 
transformation due to the slightly redder sampling of the 
$F702W_{z=0.4}$ filter with respect to the $J_{z=0}$ filter,
we still include less Irr and more E/S0 
when selecting the galaxies in the $F702W_{z=0.4}$ band than we would when 
observing the Cl0939+4713 galaxies exactly in the $J$ restframe.  We quantify 
this (negligible) effect in Sect. 4.

\titleb{Pass-band: classification}

The morphological classifications in Coma and Cl0939+4713 are not performed 
in the same restframe passband, and therefore there could be systematic
differences in the morphological classification because of the 
passband dependence of the contrast of spiral arms
and dust with respect to the galaxy main body.

However, such systematic differences apply only to a minority of galaxies, 
those whose type is on the borderline between two classes, and we verified 
that the passband dependence of the morphological type, at the sampled 
wavelengths, is negligible for galaxies at low redshift.

\titleb{Definition of the used Hubble types}

It is clear that, whatever definition of the Hubble types is adopted, it must 
be the same for distant and for nearby galaxies, which implies that the method 
used for classifying the galaxies has to be the same in both cases, although 
this requirement is not always satisfied in the literature (see the 
Introduction). 

The method introduced by Michard \& Marchal (1993), which allows one to 
assign Hubble types to galaxies according to the presence or absence of 
morphological components, such as bar, disk, bulge, halo, dust or spiral 
arms, gives reproducible morphological estimates and has the advantage that it 
has already been applied to galaxies of the local Universe (Michard \& Marchal 
1994 and reference therein, Andreon 1996b), 
and in particular to our two Coma samples (Andreon et al. 1996a,b). Therefore 
we decided to use this method to classify the galaxies of Cl0939+4713 as well. 

In this scheme, ellipticals are galaxies with elliptical isophotes and almost
linear surface brightness profiles in the $\mu$ {\it vs.} $r^{1/4}$ plane,
lenticulars have elliptical isophotes, but their major axis surface
brightness profile presents a bump, and, finally, spirals do not have
elliptical isophotes.  Galaxies not following these typical trends are not 
classified. Details can be found in Michard \& Marchal (1993) and Andreon \& 
Davoust (1996).

\titleb{Membership}

It is important to remove background or foreground galaxies from the sample
when we are interested in the properties of the galaxy population
in a selected region of the Universe.

At small distances, galaxy redshifts are available for most samples, and 
one can therefore reject interlopers on the basis of their velocity relative 
to the cluster center. We did this for the two samples in Coma (Andreon
1996a and Andreon et al. 1996a).

In Cl0939+4713, redshifts are available for only a small number
of galaxies (25 out of 83 brighter than $F702W=22.2$ mag in the field of view),
and  the membership of galaxies must be determined in a different way.

In this field, we statistically expect eleven background or foreground 
galaxies brighter than $F702W=22.2$ mag from the counts-magnitude relation of 
Smail et al. (1995), whose integral is normalized at $F702W=27$ mag as in 
Seitz et al. (1996).  We have identified nine of them with certainty (DG 234, 
DG 349, DG 396, DG 398, DG 403, DG 407, DG 420, DG 440 from their redshifts 
and DG 380 because it is interacting with DG 398, an interloper).  Three other 
galaxies (DG 292, DG 384 and DG 385) are suspected of being background quasars 
(see Sect. 4).  We have therefore identified 12 probable (most of them 
certain) background or foreground galaxies; since 11 were expected, it is 
reasonable to assume that all the other galaxies in the field of view 
(brighter than $F702W=22.2$) are cluster members. 
We have identified almost all non member galaxies, in spite of the
small fraction of galaxies with known spectrum (1/3), because DG 
chose to take spectra of blue galaxies.

We can confirm in an independent statistical way that the sample is not 
significantly contaminated by interlopers, using spectrophotometric redshifts 
taken from Belloni \& R\"oser (1996), as if they were redshifts based on 
emission or absorption lines. The spectrophotometric redshifts exclude seven 
galaxies from our sample, but do not exclude seven others which are 
interlopers by their kinematic redshift. 

Therefore, the resulting sample is likely to be an absolute magnitude complete
sample of galaxies, like the two Coma ones. 

\titlea{The morphological types of galaxies in Cl0939+4713}

In order to exclude from the sample galaxies partially outside the field
of view, the studied field for Cl0939+4713 is $\sim2$ arcsec smaller on both
sides of the field of view of each of the three CCDs that compose the
Wide Field Camera 2. 

With the images of Cl0939+4713 at our disposal, we classify
galaxies down to $F702W=22.2$ mag, corresponding to $M_J=-19.0$
mag, which we conservatively take as our magnitude limit for
morphological classification. 

Such faint galaxies have not yet been morphologically classified in the Coma 
cluster, the deeper samples reaching  $M_J=-19.3$ mag. Therefore, in the 
following, when we make comparisons among these samples, we limit the deeper 
one (Cl0939+4713) to the magnitude of the shallower one. 

Table 1\fonote{Table 1 is presented only in electronic form} presents a
concise description of all galaxies brighter than $F702W=22.2$ mag in the
{\it HST} studied field of view with the exclusion of certain
background/foreground galaxies. We present the morphological type (column
14), a characterization of the galaxy envelope (column 13) and many other
geometrical parameters, detailed in the legend of Table 1,
together with notes to the individual galaxies. 

We did not succeed in attributing a morphological type
to only 8 galaxies out of 71. 

Three of them are morphologically unresolved galaxies, i.e. resolved sources 
but not large enough to be assigned a morphological type. These galaxies have 
small radii for their magnitude. They are, from the 
morphological point of view, very similar to the quasar DG 440 at $z\sim2$, 
present in our field, which suggests that they could be background quasars. 
For this reason we exclude them from the sample. This decision does not affect 
our comparison of Hubble types since these galaxies are certainly not 
ellipticals, lenticulars or spirals. This decision affects only marginally our 
comparison of the whole population of galaxies in Cl0939+4713 with the Coma 
cluster one, because, if these galaxies do belong to Cl0939+4713, one is 
fainter than the absolute magnitude limit of the Coma-center sample, and three are 
fainter than that of the Coma-whole samples.  They are therefore already 
excluded from the comparison sample because they are too faint. 

Five other galaxies are classified as peculiar. Three of them (DG 311, DG 318
and DG 436) present features typical of interacting galaxies, such as
tidal arms. Among these three galaxies, the brightest cluster galaxy (DG
311) shows a tidal arm of very low contrast (and unusual $g-r$ and $r-i$
colors), 
but a normal surface brightness profile and
elliptical shape (at small radii). The two other galaxies (DG 137 and DG
339) present pear-shaped isophotes and a detailed morphology that we did
not find in any Coma galaxy, but again one is fainter than the absolute
magnitude limit of the Coma-center sample, and both are fainter than that
of the Coma-whole sample.

\titleb{Similarity with present day galaxies}

The fact that a Hubble type can be assigned to almost
all galaxies of Cl0939+4713, 
even with classification criteria as strict as ours, reinforces DOBG's 
impression of the ordinariness of galaxies of Cl0939+4713. First of all, the 
galaxies with elliptical isophotes have de Vaucouleurs-like surface brightness 
profiles, as present day ellipticals, or a bump along the major axis, as 
lenticulars. When this bump is present, it is always linked to a change in the 
isophote axis ratio, as for present day lenticulars.  The $e_4$\fonote{The 
$e_4$ parameter (often called $a_4$ in the literature), when positive, points 
out an excess of brightness along the major axis of the galaxy, with respect 
to a galaxy with perfect elliptical isophotes.} profile of distant lenticulars 
is similar to that of present day ones. In other words, distant and nearby 
galaxies with similar surface brightness profiles also have similar 
ellipticity and $e_4$ profiles. The similarity between lenticulars holds even 
on a statistical basis: the proportion of lenticulars of Cl0939+4713 with a 
decreasing, stationary or increasing outer ellipticity profile (which leads to 
the envelope classification of exD, thD, spH in Table 1; see Michard \& 
Marchal 1993 and Andreon 1996b), is the same as in Coma (0\%/50\%/50\%). 

Furthermore, in early-type galaxies, the more important (i.e. larger) 
coefficient measuring the deviation of the isophotes from the perfect 
elliptical shape is the $e_4$ parameter, as for present day early-type 
galaxies. 

The similarity of distant galaxies to local counterparts holds for the
shape of the broad spectrum too.  Distant and nearby early-type galaxies have 
the same spectrum. Distant spirals present a variety of spectra, as present day 
spirals. In fact, comparing our morphological types to the spectrophotometric 
types of Belloni \& R\"oser (1996), we found that, out of 36 early-type 
galaxies, 34 have the spectrum of an early-type galaxy, one that of an Sbc and 
the last one that of an E+A. Out of 29 spirals, 11 have the spectrum of an 
S/Im, 10 an early-type spectrum and 8 are E+A's. A similar conclusion has been 
reached by Belloni et al. (1995) in their comparison of their 
spectrophotometric type to DOBG's morphological types. Similarly, the break at 
4000 \AA \ (taken from the same reference) is more prominent in Es than in Ss, 
as in the local universe (Belloni \& R\"oser 1996).

\titleb{Comparison with pre- and post-refurbished types}

The comparison between DGOB's morphological type estimates and ours,
from pre- and post-refurbished {\it HST} images respectively, shows that
the two types are compatible, once differences in the observational
material are taken into account, for example our detection of a disk or a
spiral pattern that DGOB do not see because of the lower quality of their
data. Galaxy to galaxy comparisons are presented in the ``Notes to the
galaxies" in Table 1 for remarkable cases. 

Post-refurbished {\it HST} morphological types for 31 galaxies, estimated
by Dressler et al. (1994b) are listed in Stanford, Eisenhardt \&
Dickinson (1995).  We attribute the same morphological label to 24 (77\%) of
them. The morphological type of the other galaxies differs by one
class (i.e. no Es in one study have been classified as Ss in the other
one), and we have a tendency to classify the galaxies into a slightly
later type than Dressler et al. (1994b). The latter authors present a
similar rate of agreement among themselves. The rate of agreement between our 
types and Dressler et al.'s (1994b) for Cl0939+4713 is the same as that 
between our types and those from other authors for Coma galaxies (Andreon \& 
Davoust 1996), including Dressler. 

Please note that, since our method for estimating the Hubble type is the
same for both nearby and distant galaxies and it is highly reproducible,
this comparison does not measure the reproducibility of our Hubble type
estimates but only how similar our classification system is to the other
one.

\titleb{Morphological composition of Cl0939+4713}

\begtabfull
\tabcap{2}{Morphological composition of Coma and Cl0939+4713
for various cluster regions and magnitude limits}
\halign{#\hfill&\quad\hfill#&\quad\hfill#&\quad\hfill#&\quad\hfill#&
\quad\hfill#\cr
\noalign{\smallskip\hrule\medskip}
Region & $N_{gal}$ & \multispan4{\hfill Percentage of \hfill}\cr
& & \multispan4{\hrulefill}\cr
& & E & S0 & S & other\cr
\noalign{\medskip\hrule\medskip}
\multispan6{\hfill $M_J=-19.0$ \quad \hfill}\cr
Cl0939+4713& 71& 14& 37& 42& 7\cr
\noalign{\medskip}
\multispan6{\hfill $M_J=-19.3$ \quad \hfill}\cr
Cl0939+4713& 67& 15& 34& 43& 7\cr
Coma-center& 96& 23& 52& 19& 6\cr
\noalign{\medskip}
\multispan6{\hfill $M_J=-19.8$ \quad \hfill}\cr
Cl0939+4713& 52& 13& 31& 48& 7\cr            
Coma-center& 71&  27& 48& 21& 4\cr
Coma-whole&  186& 25& 42& 31& 2\cr
(Coma-center)$^C$ $\cap$ (Coma-whole)& 115& 24& 39& 37& -\cr 
\noalign{\medskip\hrule\medskip}}
\noindent Note: (Coma-center)$^C$ is the complementary of the Coma-center
sample.
\endtab

Table 2 presents the morphological composition of Cl0939+4713 in the studied 
region down to our limiting magnitude for morphological classification ($M_J=-
19.0$ mag) and down to the adopted limits for galaxy classification of two 
comparison samples. Note that the area surveyed by the field of view of the 
{\it HST} has a radius which is half that of the Coma-center sample. 

Down to both common absolute magnitudes, the spiral fraction of Cl0939+4713 is 
$\sim20$\% higher than in Coma, and lenticulars are less abundant by 10 to 20 
\%. Ellipticals are $\sim10$\% less abundant in Cl0939+4713 than in Coma. 

The fact that the $F702W$ filter does not exactly match the
redshifted $J$ galaxy emission has a negligible influence, since
our numbers of galaxies change by 1 on the average when
differentially shifting by 0.1 mag the luminosity function of one or two
morphological types.

DOBG compare the morphological composition of Cl0939+4713
to the ``typical" one of rich clusters and of the field.
They find a larger difference than we do between the spiral composition of 
Cl0939+4713 and that of nearby clusters.  But their estimate of the 
spiral fraction in nearby clusters is smaller than what we find in Coma;
this is expected because of the general tendency of morphologists to 
underestimate the spiral fraction of nearby clusters, which we have shown to 
exist in Coma, Perseus and other nearby clusters (Andreon 1993, 1994, 1996a, 
Andreon \& Davoust 1996). The main reason for the disagreement between 
DOGB and us is that the images used by morphologists to classify galaxies in 
nearby rich clusters generally have a lower resolution than that necessary for 
a well controlled comparison, as explained in Sect. 3.1. 
Also, DOBG compare the morphological fraction of Cl0939+4713 down to $M_J 
\sim -18.1$ mag to that of nearby clusters down to an unspecified absolute 
magnitude. Since no morphological types have been published for such faint 
galaxies in rich nearby clusters, their comparison sample must be limited to a 
brighter absolute magnitude, thus invalid, unless they used unpublished 
observations of a deep sample. 

From the inspection of Figure 5 of DOBG, we note that many faint galaxies
in Cl0939+4713 are spirals.  Therefore, using a brighter limiting
magnitude for the Cl0939+4713 galaxies, similar to ours and to that
used for rich nearby clusters, DOBG would probably find a lower
proportion of spirals in Cl0939+4713, comparable to ours, and closer to
the one estimated for nearby clusters. 
At any rate, they agree with us on the spiral fraction for the 31 galaxies in 
common.

Therefore, the morphological composition of Cl0939+4713 does indeed differ 
from the one of nearby clusters, but {\it by a factor two less} than what is 
found by DOBG. The necessity of a morphological evolution of galaxies in 
clusters is thus strongly reduced. 
This is discussed in Sect. 7.3.

\titlea{The properties of the Hubble types}

Galaxies which are morphologically similar are not necessarily similar in 
their physical properties, especially when they have different ages
(i.e. when they are at different redshifts).  In other words, the similarity
of morphological composition of Cl0939+4713 and Coma does not imply that
the galaxies in the two clusters have the same physical properties.
To verify this possibility, we must compare the luminosity function,
color and surface brightness distributions, morphological segregation, and
other properties of the morphological types of the two samples.

\titleb{Astrophysical parameters and method of analysis}

The following physical parameters of the Cl0939+4713 galaxies were considered. 
\newline 
-- Positions, $F702W$ isophotal magnitude, isophotal area at the 25.3 $F702W$ 
magnitude arcsec$^{-2}$ were taken from Seitz et al. (1996) for all but three 
galaxies. The $g-r$ and $r-i$ colors were taken from DG92 for all but two 
galaxies. This sample is selected in the $F702W$ filter ($\sim J$ restframe) 
and is practically 100 \% complete (for galaxies of normal central brightness, 
i.e. $J<25$ mag arcsec$^{-2}$). 
\newline 
-- The infrared magnitudes $J, H, K$ and through an intermediate filter 
centered at 7840 \AA, referred to as visual, came from Stanford, Eisenhardt \& 
Dickinson (1995) for 46 galaxies selected for having $K<18.00$ mag and in the 
{\it HST} field of view. This sample, selected in $K$, is expected to be 
$\sim90$ \% complete in magnitude down to $K=18$ mag. 
\newline 
-- The morphological types of the galaxies came from Table 1. 

The panels of Fig. 1 show the spatial distribution of all the galaxies and
for the three morphological types. Spirals are uniformly distributed, 
lenticulars are mainly found in the two upper CCDs and Es in the upper halves 
of these two upper CCDs. 

We then computed the following quantities and their distribution for each 
morphological type. 

\newline
-- the positions ($x, y$); they are Seitz et al.'s positions rotated clockwise 
by 20 degrees around the cluster center adopted by us (the dominant galaxy DG 
311), to align the $x$ and $y$ axes with the privileged directions suggested by 
the spatial distribution of the galaxy types (see Fig. 1);
\newline 
-- The the clustercentric distance and the angle with respect the
privileged direction ($\theta$);
\newline 
-- the luminosities and ($g-r, r-i, J-H, H-K, visual - K$) colors; for the
infrared colors we split the sample in only two classes, (E+S0) and S,
because of low statistics;
\newline 
-- the local density, defined as the density inside the smallest circle 
containing the 10 nearest galaxies (Dressler 1980), the distance to the
nearest neighboring galaxy ($d_1$); they were computed using 
Dressler \& Gunn's (1992) catalogue, limited to the Gunn $r$ = 22.7 mag in 
order to avoid border effects in the density and the distance to the nearest neighboring galaxy computations; 
\newline 
-- the mean surface brightness, $<$SuBr$>$; it was computed taking into 
account the galaxy ellipticity. This parameter is missing for the two galaxies 
not present in the Seitz et al. catalogue (DG 371 and DG 438, see Sect. 3.3). 

The method used to test the reality of the differences (if any) between the 
galaxy properties along the Hubble sequence has been presented and thoroughly 
tested in Andreon (1994, 1996a,b). In summary, the probability that an 
observed difference between the distributions of a given quantity for two 
types is real (or, better, one minus the confidence level for rejecting 
the hypothesis that the two classes are drawn from the same parental 
distribution) is computed by 100000 Monte Carlo simulations in which 
the galaxy types are shuffled randomly. As the number of objects in each class 
is small, it is important to use robust statistics. We used robust estimates 
of the distribution moments, even if, for the sake of clarity, we referred to 
them by their un-robust names (for example the central location index is 
called mean).  

\begtabfullwid 
\tabcap{3}{Probability, expressed in percentage, that two
galaxy classes c1 and c2 have the same parental distribution in distance
$y$, azimuthal distance $\theta$, color $g-r, r-i$, mean surface
brightness $<$SuBr$>$.  Probabilities of other parameters were also tested,
but gave null results.  Only pairs of morphological types having less than
0.2 \% probability to be drawn from the same parental distribution are
listed. For the sake of clarity, probabilities larger than 15 \% are
replaced with blanks.  Null results are not listed but are discussed in
the text when interesting.  The probabilities T, F, TI and SI refer to
differences in the mean, dispersion, kurtosis and skewness (the names are
those of the classical tests). The probabilities KS and P5 refer to the
classical Kolmogorov-Smirnov test and to the vector (T, F, TI, SI, KS) (see
Andreon 1994 and 1996a,b for details).}
\halign{#\hfill&#\hfill&\quad\hfill#\hfill&\quad#\hfill&\quad#\hfill&
\quad\hfill#&\quad\hfill#&\quad#\hfill&\quad#\hfill\cr
\noalign{\medskip\hrule\medskip}
 c1 & c2 &  quantity &  \hfill P5  & \hfill  T  &  \hfill F  &   TI \hfill&  \hfill SI  &  \hfill KS   \cr
\noalign{\medskip\hrule\medskip}
 E  & S0 &  $y$    &  0.175&  5.050&       & 2.548&        &  4.952 \cr
 E  &  S &  $y$    &  0.009&       &  5.253& 0.209&        &  3.854 \cr
 E  &  S & $\theta$&  0.147&       &       &11.393&        &        \cr
 E  &  S &  $g-r$  &  0.000&  0.010&  0.001&      &        &  0.000 \cr
 S0 &  S &  $g-r$  &  0.000&  0.000&  0.000&      &        &  0.000 \cr
 E  &  S &  $r-i$  &  0.001&  3.108&  0.150&      &        &  0.020 \cr
 S0 &  S &  $r-i$  &  0.000&  0.749&  2.191&      &   6.667&  0.029 \cr
 E  & S0 & $<$SuBr$>$ & 0.109& 0.125&      &      &        &  0.889 \cr
 S0 &  S & $<$SuBr$>$ & 0.007& 0.228&      &      &        &  0.768 \cr
\noalign{\medskip\hrule\medskip}
}\endtab

\titleb {$J$ Luminosity function}

The global luminosity function of Cl0939+4713 has a maximum at $M_J=-20$
mag (Figure 2a, left panel), which is not due to incompleteness of the
catalogue, since the limiting magnitude is at least four magnitudes
fainter.  Our statistical tests show that the global luminosity function
of Cl0939+4713 pretty well matches that of the whole cluster down to
$M_J=-19.8$ mag and that of Coma's core down to $M_J=-19.3$ mag, our
limiting magnitudes (Figure 2a, top-center and top-right panels, and note
the absence of any row containing $L$ in Table 3, see the legend of Table
3). The 68 \% confidence limits for the differential distance moduli
between Coma and Cl0939+4713, inferred from matching the respective
luminosity functions are +6.0 and +6.6 mag; they do not exclude any
reasonable luminosity evolution, such as the one expected by passive
evolution, or value of $q_0$, since the expected differential distance
modulus, for $q_0=0.5$, is +6.4 mag. 

We do not detect any statistically significant difference between the 
luminosity functions of the different morphological types of Cl0939+4713 (and 
there is no line for such a quantity in Table 3); 
this is in agreement with what we found for a sample of galaxies of 
approximately the same size in Perseus (brighter than $M_V=-19.5$ mag, Andreon 
1994), but at variance with what we found for a larger sample of galaxies in 
Coma (brighter than $M_J=-19.8$ mag). 

The statistical significance of the difference between the luminosity
functions of the Cl0939+4713 and Coma types (Es+S0s and Ss) is smaller
than $2\sigma$ (see also Figure 2a, center and right panels, and note the
absence of any row containing $L$ in Table 3), confirming DOBG's
suggestion. 

There is no statistical evidence, larger than $2\sigma$, for a difference
between the luminosity functions of the Cl0939+4713 and Coma types
(Es+S0s and Ss) (see also Figure 2a, center and right panels,
and note the absence of any row containing $L$ in Table 3), confirming
DOBG's suggestion.

\titleb {Mean surface brightness} 

The mean surface brightness distribution of S0s differs from those of Es
and Ss, in part because S0s are brighter in the mean (by 0.2 mag
arcsec$^{-2}$) (Figure 2b and Table 3,  
T column, last two rows).

The measured dispersions in mean surface brightness are 0.23 and 0.18 mag 
arcsec$^{-2}$ for Es and S0s respectively, and are likely to be due to 
photometric errors. That is, Es and S0s are separately homogeneous in mean 
surface brightness within the present photometric accuracy. In fact, an error 
a 0.1 mag arcsec$^{-2}$ in the determination of the galaxy isophotal 
brightness, which is possible in our measures, translates into a 0.16 mag 
arcsec$^{-2}$ error in the mean surface brightness for elliptical galaxies 
whose effective radius and surface brightness obey the Kormendy (1977) 
relation. 

In Coma, we also found that S0s have a higher surface brightness than Es by 
0.2 mag arcsec$^{-2}$. We cannot compare directly the mean surface 
brightnesses of the types in Cl0939+4713 and in Coma, since they are computed 
within different restframe isophotes. We can only say that each of the two 
classes is homogeneous in the two clusters and that 
the difference in mean surface brightness                  
between Es and S0s is the same in both clusters. The zero point for 
the comparison of surface brightnesses
at different redshifts is given 
by Pahre, Djorgovski \& de Carvalho (1996), who found that the 
early-type galaxies studied in Cl0939+4713 occupy the same locus as the Coma 
galaxies in the effective radius vs effective surface brightness plane. This
implies that Es and S0s in Cl0939+4713 have the same mean surface brightness
as the respective classes in Coma.
 
In the study of the whole Coma cluster, we found that S0s have fainter
mean surface brightnesses than Ss, contrary to our finding for Cl0939+4713.
However, we are sampling different portions of the two clusters,
and the presence of spirals with high surface brightness in Coma is
limited to large radii (since overbright spirals are blue and blue
spirals avoid the cluster center; see Andreon 1996a), not sampled in
Cl0939+4713. Sampling only the core of the Coma cluster (say within a
radius of 500 Kpc), as we do for Cl0939+4713, spirals have fainter surface
brightnesses than S0s, as in Cl0939+4713. 

The dispersion in mean surface brightness of Ss in Cl0939+4713 is larger
than for early-type galaxies (0.34 mag arcsec$^{-2}$ vs. $\sim$0.2 mag
arcsec$^{-2}$), as in Coma. 

\titleb {Optical colors}

\titlec {Early-type galaxies}

Es and S0s have the same colors (formally 1.56 and 1.52 mag in $g-r$ and 0.68 
and 0.64 mag in $r-i$; see also Figure 2b, and note the absence of any row 
involving these colors and morphological types in Table 3). In $r-i$, the 
dispersion is 0.05 and 0.09 mag for ellipticals and lenticulars respectively, 
and reasonably of the order of the photometric accuracy \fonote{Dressler \& 
Gunn (1992) do not estimate this quantity. A rough estimate is given by $\sqrt 
2$ times the photometric error in $r$, which in turn can be estimated from the 
scatter in $F702W-r$ for early-type galaxies assuming a negligible intrinsic 
scatter in the $F702W-r$ color and a negligible photometric error in $F702W$, 
two reasonable assumptions. This give $\sigma \sim 0.08$ mag}. In $g-r$, the 
dispersion around the color-magnitude relation is 0.07 and 0.08 mag for 
ellipticals and lenticulars respectively, again within the 
photometric accuracy. Therefore, both Es and S0s are homogeneous in these two 
colors, which sample the $\sim U-B$ and $\sim B-V$ restframe colors. This 
extends the homogeneity of early-type galaxies from visual and near infrared 
colors (Stanford, Eisenhardt \& Dickinson 1995) to near ultraviolet and 
blue--visual colors. 

The slope of the color-magnitude relation in both $g-r$ and $r-i$ colors (--
0.118 in $g-r$ vs. $F702W$) is compatible with the one of the Virgo cluster 
(Visvanathan \& Sandage 1977), and of other nearby clusters (Garilli et al. 
1996), extending to much bluer colors a similar finding by Stanford, 
Eisenhardt \& Dickinson (1995). 

Nothing can be said about the normalization of the color-magnitude relation 
(i.e. about the galaxy colors reduced to a fixed absolute magnitude) because 
the zero point determination of the colors in Dressler \& Gunn (1992) has a 
large error (0.1--0.2 mag). 

The $g-r$ filter samples the 4000 \AA \ break in the galaxy restframe
and is therefore sensitive to the galaxy star formation and evolutionary
history, whereas the visual and near-infrared colors are more sensitive to
metallicity. The homogeneity of the colors of early-type galaxies in
Cl0939+4713 strongly suggests a common star formation and evolutive
history, as found in the optical colors for Coma early-type galaxies by
Bower, Lucey \& Ellis (1992).  The measured dispersion in the blue color (0.08 
mag) for early-type galaxies in Cl0939+4713 is fully compatible with the 
photometric errors. Using the prescriptions of Bower, Lucey \& Ellis (1992), 
we can put a lower limit on the age of early-type galaxies, which depends on 
the degree of coordination between galaxy age and galaxy formation rate. We 
find that early-type galaxies formed at least 11 Gyrs ago if their time of 
formation was 1 Gyr or if the ratio of the spread in formation time to the 
characteristic time scale at formation is 0.3, similar to what has been found 
by Bower, Lucey \& Ellis (1992) for Coma galaxies.  However our result, which 
assumes an intrinsic scatter $\sqrt{2}$ times smaller than the observed 
scatter, makes a smaller look-back time extrapolation since galaxies in 
Cl0939+4713 are 6 Gyrs younger than present day ones. That is, the redshift of 
(early type) galaxy formation is greater than 1.5 for the adopted cosmology.

Please note that, using space based data, it is possible to measure galaxy
colors with small errors (because of the fainter sky background in space), 
allowing one to put a stronger constraint on the lower limit of the galaxy 
age, and consequently an upper limit on the value of $H_0$. 

\titlec {Spirals}

The color distribution of spiral galaxies 
markedly differs (as expected) from those of the other coarse and detailed 
types, because there are no blue ($g-r<1$ or $r-i<0.5$) early-type galaxies 
(Table 3 and Figure 2b). The dispersion in the $g-r$ and $r-i$ colors for 
spirals (0.14 and 0.15 mag respectively) is larger than the photometric 
accuracy, making these galaxies heterogeneous in colors, as the Coma spirals. 

A closer inspection of the spiral color distribution shows that the 
Cl0939+4713 spirals do not have unusual colors in general, since they have the 
K-corrected colors of present day field spirals. However we note that the 
spirals in Coma's core are red, since no blue spirals have been found there
(Andreon 1996a), and the same applies to other nearby clusters 
(Oemler 1992). Therefore the Cl0939+4713 spirals have an unusual color for 
their distance from the cluster center. This is the well known excess of blue 
galaxies first noted in the core of distant clusters by Butcher \& Oemler 
(1984). 

\titleb {$J-H$, $H-K$ and $visual-K$ infrared colors}

No differences have been found in the infrared $J-H$ and $H-K$ color 
distributions among the $K$ selected sub-samples of ellipticals, lenticulars 
and spirals (there is no line for such quantities in Table 3).
However the three types only count 6, 14, and 14 galaxies 
respectively.  No difference has even been found between the E+S0 and S color 
distributions. All these color distributions have a dispersion compatible with 
the photometric accuracy ($\sim$ 0.15 mag at $K \sim 17.5$). 

Surprisingly, the same applies to the $visual-K$ color, but for the marginal
detection of a difference between S0s and Ss and between Es+S0s and Ss,
as already noted by Stanford, Eisenhardt \& Dickinson (1995). 

\titleb {Morphological segregation}

There is no statistical
evidence of a radial, density or nearest galaxy segregation,
but there is a clear indication of the presence of a preferred direction,
along or orthogonal to which the types are more or less dispersed, as in
Coma (Andreon 1996a) and Perseus (Andreon 1994), giving a statistical
confirmation to the visual impression previously pointed out (there is no 
line for $r$, $\rho_{10}$ and $d_1$ in Table 3 but there are for $y$ and 
$\theta$; see also Figure 1). 

This is the third case, out of three studied by us, in which the primary 
source of the morphological segregation is not the radial or density 
segregation, but the presence of a privileged direction. 
In Coma, the privilegied direction 
is also present in the velocity field (Biviano et al. 1996).

For the Coma and Perseus clusters, the privileged direction is aligned
with the direction of the filament in which they are embedded (Andreon
1996a,b) and also with the position angle of the outer isophotes of the
cluster X-ray emission (note their close correspondence with the chain of
bright galaxies in Perseus and with the NGC 4938 group in Coma on the
isophote maps of Schwarz et al. (1992) and White, Briel \& Henry (1993)).
The privileged direction of the type distributions in Cl0939+4713 seems
to coincide with the major axis of the outer isophotes of the X-ray
emission, even though it is ill defined because of the irregularity of
these isophotes. 

\titlea {Summary of the results}

The results of the quantitative analysis of magnitude complete samples of
galaxies in Cl0939+4713 and in Coma, classified by the same method, using
images of very similar resolution, selecting the samples in very
similar pass-bands and limiting them at the same absolute magnitude, 
can be summarized as follows. 

1) We confirm the ordinariness of the morphology of distant early-type 
galaxies in Cl0939+4713, suggested by DOBG from a visual examination of the 
appearance of these galaxies. Galaxies of a given Hubble type are similar to 
those of Coma in their surface brightness, ellipticity and $e_4$ profiles, and 
in the relative frequencies of the outer envelope classifications. 

2) The cluster Cl0939+4713 is rich in spirals, twice as much as in Coma, but
half than previously estimated. The ratio between Es and S0s is larger in 
Cl0939+4713 than in Coma. 

3) Barred lenticular galaxies are {\it apparently} missing in Cl0939+4713. But 
the coarse sampling of the point spread function of the {\it HST} images makes 
the detection of a bar very difficult, and therefore their absence at high 
redshift have to be interpreted with caution. 

4) There are no statistical differences between the global luminosity
functions of Cl0939+4713 and Coma, both limited to $M_J=-19.8$ or
$M_J=-19.3$ mag, our present magnitude limits for the two Coma samples. The 
same applies to the luminosity function of the types, as already suggested by 
DOBG. 

5) The mean surface brightnesses of Es and S0s differ by the same amount in 
Cl0939+4713 and in Coma (S0s are 0.2 mag arcsec$^{-2}$ brighter than Es). Both 
Hubble classes are separately homogeneous in $J$ mean surface brightness. 
Thanks to Pahre, Djorgovski \& de Carvalho's (1996) finding, our result 
implies that the E and S0 classes in Cl0939+4713 have the same mean surface 
brightness as the respective classes in Coma. Lenticulars have a 
brighter mean surface brightness than spirals in the core of Cl0939+4713, as 
in Coma's core, but at variance with the spiral population outside Coma's 
core. Please note that our early-types are brighter than the ``ordinary" class 
of Capaccioli et al. (1992) and therefore that the homogeneity only concerns 
bright galaxies. 

6) The Es have the same $\sim U-B$ and $\sim B-V$ restframe colors as S0s, 
and both classes are homogeneous in these colors (once the color-magnitude 
relation is subtracted) in Cl0939+4713.  This is also true in Coma. 
These galaxies are at least 11 Gyrs old.

7) The color-magnitude relation for early-type galaxies in Cl0939+4713
has the same slope as in Coma, extending to the visible-blue and near
ultraviolet what Stanford, Eisenhardt \& Dickinson (1995) found for the visual
and near infrared colors. 

9) The Ss in Cl0939+4713 have the $\sim U-B$ and $\sim B-V$ restframe
colors of present days field Ss, but are bluer than the spirals in Coma's
core, since the latter are redder than the ones in the field and also
than the ones outside Coma's core. 

10) The morphological types in Cl0939+4713 are segregated in the same way
as in Coma and Perseus, i.e. without a clear evidence for a density or
clustercentric segregation, but revealing the presence of a privileged
direction that coincides with the position angle of the major axis of the
outer X-ray isophotes in the three clusters. 

\titlea{Conclusions}

\titleb{Es and S0s}

From all points of view, the early-type galaxies in Cl0939+4713 share the
properties of their local counterpart galaxies, from which they are
indistinguishable, at least within our photometric and statistical errors and
for the considered, mainly photometric, galaxy properties. Please note
that E+A galaxies in Cl0939+4713 morphologically are spirals and not
ellipticals (Belloni et al. 1995 and Sect. 4). 

The only difference known to date between the Cl0939+4713 early-type
galaxies and the local respective counterparts is their rest-frame $J-K$
colors (Stanford, Eisenhardt \& Dickinson 1995). However, Stanford,
Eisenhardt \& Dickinson (1995) use Bower's (1992 and private
communication) colors of Coma galaxies as a reference sample, which show a
systematic difference with respect to Recillas-Cruz et al.'s (1990) $J-K$. 

Since there are no blue early-type galaxies (at least in the studied
samples), the ancestors of early-types will first become red, which takes
$\sim 1$ Gyr (Charlot \& Silk 1994), and then will take the early-type
morphological appearance. This implies that morphological
components characterizing spiral galaxies, such as spiral arms, HII
regions and dust in large quantity, need more than 1 Gyr to be made
undetectable by whatever mechanism is operating in clusters if S galaxies
change morphological type. 

\titleb{Ss}

Given their distance from the cluster center, the spirals in Cl0939+4713 have 
unusual properties. Their frequency with respect to the other morphological 
types is higher than in present day clusters, their colors are similar to 
those of field spirals, much more than to those of spirals in the center of 
Coma-like clusters, which are red. Their spatial distribution and mean surface 
brightness are not typical of red or infalling (blue) spirals in Coma, since 
the Cl0939+4713 spirals are not overbright for their magnitude and do not 
avoid the cluster center. This suggests that, in Cl0939+4713, the blue spirals 
are just spirals with a normal star formation rate, as suggested by the 
optical colors of blue galaxies in many other distant clusters (Rakos \& 
Schombert 1995; Rakos, Maindl \& Schombert 1996). 

Why are blue spirals with a normal star formation rate present in the core of 
distant clusters, but absent in the core of nearby Coma-like clusters? 

Many authors, starting with Butcher \& Oemler (1984), suggest that some kind 
of evolution is responsible for the disappearance of this blue population, 
which becomes red or undetected. However, the fading and/or destruction of the 
distant blue population, to make them undetected in present day clusters, 
seems unreasonable (Rakos \& Schombert 1995), since most galaxies have to 
disappear. Changing morphological type seems even more unreasonable,
because of 
the homogeneity of the photometric properties of early-type galaxies
which implies that they are at least 11 Gyrs old.

In present day clusters, the lack of blue galaxies in the cluster core,
the color distribution of spiral galaxies and many of their properties
can be explained if spirals falling in clusters have a starbust due to
the ram pressure in the hot gas (Bothun \& Dressler 1986) that consumes
the galaxy's gas reservoir. During the burst, these galaxies become bluer
and brighter in mean surface brightness, then, shortly after the burst,
they become as red as ellipticals (Charlot \& Silk 1994), thus explaining
the presence of red spirals in cluster cores. Furthermore, the
existence of galaxies exhibiting spectral signatures consistent with the
presence of intermediate age stellar populations (Couch \& Sharples 1987;
Lavery \& Henry 1988, Dressler \& Gunn 1992), and the previously listed
photometric evidence for the blue starburst spirals in Coma, support this
scenario.  In a simplified view, during the galaxy's motion through the
intracluster hot gas, the galaxy's gas is removed by ram pressure (Gunn \&
Gott 1972) and therefore its star formation is truncated, thus accounting
for the lack of blue galaxies in the cluster core. 

This scenario can also account for the presence of many blue galaxies in the 
core of distant clusters if the conditions for galaxy gas stripping are not 
satisfied, for example if the intergalactic gas density is lower than the 
threshold for this stripping. 

X-ray observations show that the gas density distribution in Cl0939+4713 
(Schindler \& Wambsganss 1996) is quite different from that in Coma, and the 
conditions for triggering the starburst are perhaps not realized. In the 
distant cluster 3C295, which instead possesses a low fraction of blue galaxies 
(Butcher \& Oemler 1984), the gas density is high (Henry et al. 1979). 
Furthermore, the existence of nearby clusters with a high fraction of blue 
galaxies (e.g. Abell 1367) and of distant clusters with a low fraction of blue 
spirals (e.g. Cl0016+16), even in the sample studied by Butcher \& Oemler 
(1984) suggests that our present interpretation of the excess of blue galaxies 
in distant clusters is a possible alternative to the evolutive interpretation 
of the Butcher-Oemler effect. 

In order to test this working hypothesis, we are now directly measuring the
cluster gas brightness (which is a measure of the gas density) in a
uniform way for a sample of clusters, nearby as well as distant, for
which an accurate and uniform measure of the blue fraction of galaxies is 
available. 

\titleb {Morphological segregation}

Sandage (1990) suggested testing whether the segregation is the result of
a biased galaxy formation plus a current infall by measuring an evolution
of the morphological segregation. In such a case, ellipticals form in
high density regions whereas spirals in lower ones, which more recently
detached from the Hubble flow and fall in the cluster. Therefore
distant (younger) clusters have to be richer in ellipticals in their core
than present day clusters, since their outer shells, richer in late type
galaxies, have not yet fallen in. 

However we have shown, in the detailed study of three cluster (Coma, Perseus 
and Cl0939+4713), that the galaxies are more strongly segregated with respect 
to a privileged direction than in density or clustercentric distance, making a 
simple verification of the above prediction more difficult. Furthermore, the 
core of Cl0939+4713 is poorer in early-type galaxies than Coma, showing that a 
more complex scenario is needed to explain the evolution of the galaxy 
populations in clusters. 

A viable alternative for explaining why many distant clusters have higher
spiral fractions than nearby ones is to remove the assumption that the
former (or at least the observed ones) are the ancestors of the latter.
This is reasonable, since it seems that distant and nearby clusters
differ in richness and gas content. Therefore, the galaxy populations of
distant clusters do not have to evolve necessarily into those of nearby ones.
This resolves the problems encountered by evolutionary scenarios in 
accounting for the differences in galaxy populations observed in clusters 
at different redshifts.

\acknow{We thank E. Salvador-Sol\'e's for valuable comments which helped us 
clarify the discussion, and G. Longo for carefully reading the manuscript. 
We are indebted to S. van der Bergh for pointing out to us the morphological 
difference between bars in lenticulars and in spirals.  We thank J.-P. Kneib 
and P. Belloni for releasing Seitz et al.'s (1996) and Belloni \& R\"oser's 
(1996) Cl0939 galaxy catalogues in advance of publication. We also thank D. 
Bottini. This research has made use of the NASA/IPAC Extragalactic Database 
(NED) which is operated by the Jet Propulsion Laboratory, California Institute 
of Technology, under contract with the National Aeronautics and Space 
Administration.}

\begref{References}
\ref Abraham R., Tanvir N., Santiago B. et al. 1996, MNRAS 279, L47
\ref Andreon S., 1993, A\&A 276, L17
\ref Andreon S., 1994, A\&A 284, 801 
\ref Andreon S. 1996a, A\&A 314, 763 
\ref Andreon S. 1996b, PhD thesis, Universit\'e P. Sabatier, Toulouse, France %
\ref Andreon S., Davoust E. 1996, A\&A, in press 
(astro-ph/9610235)  
\ref Andreon S., Davoust E., Poulain P., 1996a, A\&AS, submitted
\ref Andreon S., Davoust E., Michard R. et al. 1996b, A\&AS, 116, 429 
\ref Arag\'on-Salamanca A., Ellis R., Couch W., Carter D. 1993, MNRAS 262, 764
\ref Barrientos L.F., Shade D., L\'op\`ez-Cruz O., 1996, ApJ 460, L89
\ref Belloni P., Bruzual A., Thimm G., R\"oser H.-J., 1995, A\&A 297, 61
\ref Belloni P., R\"oser H.-J., 1996, A\&AS, 118, 65       
\ref Biviano A., Durret F., Gerbal D., et al., 1996, A\&A 311, 95
\ref Bothun G., Dressler A., 1986, ApJ 301, 57
\ref Bower R., Lucey J., Ellis R., 1992, MNRAS 254, 589
\ref Butcher H., Oemler A., 1984, ApJ 285, 426
\ref Butcher H., Oemler A., 1984, ApJS  57, 665
\ref Capaccioli M., Caon N., D'Onofrio M., 1992, MNRAS 259, 323
\ref Casertano S, Ratnatunga K., Griffits R. et al., 1995, ApJ 453, 599
\ref Charlot S., Silk J., 1994, ApJ 432, 453
\ref Coleman G., Wu C., Weedman D., 1980, ApJS 43, 393
\ref Couch W., Sharples R. 1987, MNRAS 229, 423
\ref Courteau S., de Jong R., Broeils A., 1996, ApJ, L73
\ref Dressler A., 1980, ApJ 236, 351
\ref Dressler A., Gunn J. 1992, ApJS 78, 1
\ref Dressler A., Oemler A., Butcher H., Gunn J., 1994a, ApJ 430, 107
\ref Dressler A., Oemler A., Sparks W., Lucas R., 1994b, ApJ 435, L23
\ref Driver S., Windhorst R., Griffits R., 1995, ApJ 453, 48   
\ref Ellis R., 1996, in {\it Unsolved Problems in Astrophysics}, eds.
        J. Ostriker and J. Bahcall, Princeton, in press
\ref Fukugita M., Shimasaku K., Ichikawa T., 1995, PASP 107, 945
\ref Garilli B., Bottini D., Maccagni D., et al., 1996, ApJS 105, 191
\ref Glazebrook K, Ellis R., Santiago B., Griffits R., 1995, MNRAS, 275, L19
\ref Godwin J., Metcalfe N., Peach J., 1983, MNRAS 202, 113
\ref Gunn J., Gott J., 1972, ApJ 176, 1
\ref Henry J., Branduardi G., Fabricant D., et al., 1979, ApJ 234, L15
\ref Holtzman J.,Burrows C., Casertano S. et al. 1995, PASP 107, 1065
\ref Hubble E., 1936, {\it The Real of the Nebulae},  New Haven: Yale 
         University Press
\ref Kormendy J., 1977, ApJ 218, 333
\ref Lavery R., Pierce M., McClure R., 1992, AJ 104, 6
\ref Lavery R., Henry J., 1988, ApJ 330, 596
\ref Michard R., Marchal J. 1993, A\&AS 98, 29
\ref Michard R., Marchal J. 1994, A\&A 105, 481
\ref Oemler A., 1992, in {\it Cluster and Supercluster of Galaxies}, eds. A. 
         Fabian, (Kluwer Academic Publishers)
\ref Pahre M., Djorgovski S., de Carvalho R., 1996, ApJL 456, 79
\ref Poulain P., Nieto J.-L., Davoust E., 1992, A\&AS 95, 129   
\ref Rakos K, Schombert J., 1995, ApJ 439, 47
\ref Rakos K., Maindl T., Schombert J., ApJ, 1996, 466, 122    
\ref Recillas-Cruz E., Carrasco L., Serrano A. et al., 1990, A\&A 229, 64
\ref Saglia R., Bender R., Dressler A., 1993, A\&A 279, 77
\ref Sandage A., 1961, {\it The Hubble Atlas of Galaxies}, 
        (Washinton: Carnegie Institution)
\ref Sandage A., 1990, in {\it Clusters of Galaxies}, eds. W. Oegerle, M. Fitchett
       \& L. Danly (Cambridge University Press)
\ref Schindler S., Wambsganss J., 1996, A\&A, 313, 113
\ref Schwarz R., Edge A., Voges W., et al. 1992, A\&A 256, L11
\ref Seitz C., Kneib J.-P., Schneider P., Seitz S. 1996, A\&A, 314, 707
\ref Smail I., Hogg D., Yan L., Cohen J. 1995, ApJ 449, L105
\ref Stanford S., Eisenhardt P., Dickinson M., 1995, ApJ 450, 512
\ref Thompson L., 1986, ApJ 306, 384
\ref van den Bergh S.,, Abraham R., Ellis R., et al. 1996, AJ, 112, 359
\ref de Vaucouleurs G., 1959, {\it Handbuch der Physik}, vol. 53,
        (Berlin: Springer Verlag)
\ref de Vaucouleurs G., Buta R., 1980, ApJS 44, 451 
\ref Visvanathan N., Sandage A., 1977, ApJ 216, 214
\ref Zepf S., 1996, in {\it Sant'Agata Conference on Interacting 
       Galaxies: in Pairs, Groups, and Clusters}, eds. G. Longo, M. Capaccioli,
       Busarello (Kluwer Academic Publishers)
\ref White S., Briel U., Henry J., 1993, NMRAS 261, L8

\endref

\vfill\eject  
\null\vfill\eject  
{\bf Legend and notes of Table 1}

\noindent Content of Table 1
(1) No in the Dressler \& Gunn (1992) catalogue.

(2) logarithm of the approximate effective radius {\bf in arcsec}.

(3) Photometric evidence for a disk, coded as:
	. st strong
	. cl clear
	. ft faint
	. no none.

(4) Typical axis ratio, either its minimum value, if clearly defined,
or its value at the effective isophote in other cases.

(5) Location where the typical axis ratio  
was estimated, coded as ex for the
extremum, re for the effective isophote, 
co if the value is the same at both locations.

(6) Typical $e_4$ parameter, either its extremum value, if clearly defined,
or its value at the effective isophote in other cases. The estimates
are in \%.

(7) Location where the $e_4$ parameter was estimated, coded as ex for the
extremum, re for the effective isophote, co if the value is the same at both
locations. 

(8) Axis ratio in the envelope, i.e. at the isophote $F702W \sim 25$
mag arcsec$^{-2}$.

(9) Amplitude of isophotal twist in the range of reliable measurements,
in degrees. A dash means that the twist is difficult to be measured, 
as the isophotes are nearly circular. 

(10) Detection of a bar, coded as follows: bar (bar seen), bar?  (bar 
suspected), -no (no bar seen). 

(11) Detection and classification of a disk, coded as follows:
        emDi  (embedded disk),
	miDi  (mixed disk),
	exDi  (extended disk),
	-?Di  (detected but unclassified disk),
	-no-  (no disk seen).

(12) Detection of a spiral pattern, coded as follows:
	spiP  (spiral pattern seen),
	spiP? (spiral pattern suspected),
	-no-  (no spiral pattern seen).

(13) Classification of an envelope, coded as follows:
	spH   (spheroidal halo),
	thD   (thick disk),
	exD   (extended disk),
	pec   (peculiar envelope),
	-?-   (unclassified envelope).

(14) Our morphological classification, coded as follows:
	boE   (boxy E),
	unE   (undetermined E),
	diE   (disky E),
	SA0, SAB0, SB0, Sa, ... (as usual),
	S   (spiral of unknown stage), SB.. (barred spiral of unknown stage).

(15) An asterisk refers to notes about specific features such as important 
dust pattern, ring or lens, low SuBr, $f_4$-asymmetry, etc. and about 
uncertainties of various sources.

NB: When parameters have not been measured or
specific morphological components have not been studied, the relevant 
codes are replaced by blanks.

\noindent
Notes to galaxies:\hfill \break
     137 roundish galaxy, pear-shaped at intermediate radii, 
large twist in the envelope. SA0 galaxy with dust unresolved?\hfill \break
     138 very irregular isophotes.\hfill \break
     150 asymmetric in the outer regions.\hfill \break
     190 irregular isophotes.\hfill \break
     191 warp.\hfill \break
     211 probably barred.\hfill \break
     215 possible error of Dressler et al. (1994) in {\it HST} WF-PC1 
identification.\hfill \break
     225 irregular isophotes, HII regions.\hfill \break
     269 uncertain type, because near the edge of the image.\hfill \break
     292 more concentrated than DG 385.\hfill \break
     293 arms and HII regions.\hfill \break
     294 HII regions.\hfill \break
     299 outer irregular isophotes.\hfill \break
     302 difficult case, $r^{1/4}$ SuBr profile, boxy if anything.\hfill \break
     310 various clumps superposed.\hfill \break
     311 interacting, dust, ring.\hfill \break
     312 arms and HII regions.\hfill \break
     316 arms and HII regions.\hfill \break
     318 within the halo of DG 311 - tidal extension.\hfill \break
     320 arms, HII regions and dust.\hfill \break
     329 HII regions, dust.\hfill \break
     339 pear-shaped isophotes ($e3,f3 \ne 0$) at the $r_e$. \hfill \break
     348 arms - beautiful spiral, HII regions.\hfill \break
     360 edge-on, warped.\hfill \break
     363 tail at low subr, difficult classification due to the smallness of 
the galaxy.\hfill \break
     365 pear-shaped at low SuBr, nearby straight object (arclet?) and a 
galaxy at 1".\hfill \break
     366 boxy bulge, $f_4$ asymmetry.\hfill \break
     367 boxy inside 1 arcsec.\hfill \break
     371 arm, HII regions.\hfill \break
     375 strange companion galaxy(?). \hfill \break
     377 larger component of pair probably unresolved from WF1.\hfill \break
     383 arms.\hfill \break
     384 S? irregular isophotes - dust?\hfill \break
     385 $r^{1/4}$ profile, unresolved.\hfill \break
     393 near the edge of the image. \hfill \break
     397 superposed stars probably explain WF1 classification.\hfill \break
     400 roundish galaxy.\hfill \break
     408 slight evidence for a disk. \hfill \break
     409 elongated galaxy with flat profile.\hfill \break
     413 roundish galaxy with $r^{1/4}$ profile, irregular outer isophotes, 
and constant positive $e_4$.\hfill \break
     422 non-concentric isophotes, tidal extension, possibly interacting with 
DG 439.\hfill \break
     431 the companion galaxy DG 436 shows disturbances.\hfill \break
     432 $r^{1/4}$ profile but asym. isophotes similar to the  GMP 1646 
irregular galaxy in Coma, difficult classification.\hfill \break
     434 irregular asymmetric.\hfill \break
     436 possibly interacting with DG 431, tidal extension, irregular
isophotes,  halo offcentered. \hfill \break
     439 arms and dust.\hfill \break
     441 edge-on, type uncertain because small asymmetry points toward S.\hfill \break
     448 faint galaxy.\hfill \break
     451 HII regions, possibly interacting with DG 458 because common envelope.\hfill \break
     454 asymmetric.\hfill \break
     457 face-on, two-component galaxy (bulge+disk).\hfill \break
     458 roundish object, possibly interacting with DG 451.\hfill \break
     462 irregular isophotes.\hfill \break
     471 dust.  \hfill \break

\null
\vfill\eject
\tabcap{1}{Photometric parameters and classification of the galaxies in the sample}
{\baselineskip=9pt
\halign{\hfill#\quad&\hfill#\quad&\hfill#\quad&\hfill#\quad&\hfill#\quad&\hfill#\quad&\hfill#\quad&\hfill#\quad&
\hfill#\quad&\hfill#\quad&\hfill#\quad&\hfill#\quad&\hfill#\quad&#\hfill&\hfill#\quad\cr
\noalign{\medskip\hrule\medskip}
(1)& (2)& (3)& (4)& (5)& (6)& (7)& (8)& (9)& (10)& (11)& (12)& (13)& (14)& (15)\cr
\noalign{\medskip}
 135 &        &      &      &    &      &    &      &     &      &      & spiP &      & S          &          \cr
 137 &   -0.5 &   no & 0.98 & re &  0.0 & re & 0.80 &  30 & -no  & -no- & -no  & --  & Pec   & *        \cr
 138 &        &      &      &    &      &    &      &     &      &      & --  &      & Irr        & *        \cr
 141 &        &      &      &    &      &    &      &     &      &      &      &      & Sd         &          \cr
 150 &        &      &      &    &      &    &      &     &      &      & spiP &      & S          & *        \cr
 190 &        &      &      &    &      &    &      &     &      &      & spiP &      & S          & *        \cr
 191 &   -0.5 &   st & 0.63 & re &  7.3 & ex & 0.44 &   7 & -no  & miDi & -no  & thD  & SA0        & *        \cr
 215 &   -- &   st & 0.50 & ex & -2.0 & ex & 0.60 &   7 & -no  & -- & -no  & spH  & SA0        & *        \cr
 225 &        &      &      &    &      &    &      &     &      &      &      &      & S          & *        \cr
 231 &        &      &      &    &      &    &      &     &      &      &      &      & Irr        &          \cr
 246 &   -0.4 &   st & 0.62 & ex &  3.0 & ex & 0.63 &  10 & bar? & miDi & -no  & thD  & SAB0       &          \cr
 250 &   -0.7 &   cl & 0.80 & ex &  0.0 & re & 0.85 &   5 & -no  & -?Di & -no  & thD  & SA0        &          \cr
 269 &   -0.4 &   no & 0.90 & co &  0.0 & co & 0.96 & --  & -no  & -no- & -no  & --  & unE        & *        \cr
 292 &        &      &      &    &      &    &      &     &      &      &      &      & unresolved & *        \cr 
 293 &        &      &      &    &      &    &      &     &      &      & spiP &      & S          & *        \cr
 294 &        &      &      &    &      &    &      &     &      &      & spiP &      & Irr        & *        \cr
 299 &   -0.5 &   st & 0.56 & ex &  4.9 & ex & 0.59 &  35 & --   &      & -no  & -?-  & SB..       & *        \cr
 302 &   -- &   cl & 0.72 & co &  1.0 & co & 0.90 &  10 & -no  & emDi & -no  & spH  & SA0        & *        \cr
 303 &   -0.4 &   st & 0.56 & re &  2.3 & re & 0.60 &  20 & -no  & miDi & -no  & thD  & SA0        &          \cr
 310 &   -0.3 &   cl & 0.68 & co &  1.0 & re & 0.80 &   5 & -no  & emDi & -no  & spH  & SA0        & *        \cr
 311 &    0.3 &   cl & 0.77 & re & -2.6 & ex & 0.60 &  12 & -no  &      & -no  & -?-  & Pec      & *        \cr
 312 &        &      &      &    &      &    &      &     &      &      &      &      & S          & *        \cr
 314 &   -0.6 &   ft & 0.79 & re &  0.5 & re & 0.85 &  21 & bar? & emDi & -no  & thD  & SAB0       &          \cr
 316 &        &      &      &    &      &    &      &     &      &      &      &      & S          & *        \cr
 318 &   -0.3 &   cl & 0.97 & co &  0.3 & re & 0.88 &  50 & -no  & exDi & -no  & -?-  & Pec      & *        \cr
 319 &   -0.6 &   ft & 0.75 & co &  0.0 & co & 0.81 &  80 & bar  & -- & -no  & -- & SB0        &          \cr
 320 &        &      &      &    &      &    &      &     &      &      &      &      & S          & *        \cr
 321 &   -0.6 &   st & 0.40 & ex &  6.6 & ex & 0.58 &   2 & -no  & -?Di & -no  & spH  & SA0        &          \cr
 322 &   -0.7 &   ft & 0.71 & ex &  4.5 & ex & 0.79 &  12 & -no  & miDi & -no  & thD  & SA0        &          \cr
 329 &        &      &      &    &      &    &      &     &      &      &      &      & S          & *        \cr
 336 &   -- &   st & 0.48 & -- &  2.0 & -- & 0.71 &  15 & --  &      & -no  & -?-  & Irr        &          \cr
 338 &   -0.7 &   st & 0.45 & ex &  5.9 & ex & 0.50 &   0 & -no  & miDi & -no  & thD  & SA0        &          \cr
 339 &   -0.2 &   no & 0.80 & re & -2.2 & co & 0.75 &  12 & -no  & -no- & -no  & --  & Pec       & *        \cr
 348 &        &      &      &    &      &    &      &     &      &      &      &      & S          & *        \cr
 356 &   -0.8 &   st & 0.58 & ex &  3.3 & co & 0.83 &   5 & -no  & emDi & -no  & spH  & SA0        &          \cr
 360 &   -0.3 &   cl & 0.25 & ex & 10.3 & ex & 0.28 &   7 &      &      &      & spH  & S          & *        \cr
 361 &   -0.6 &   st & 0.55 & ex &  2.6 & ex & 0.65 &   9 & -no  & miDi & -no  & thD  & SA0        &          \cr
 363 &   -0.7 &   st & 0.62 & ex &  0.0 & ex & 0.75 &   0 & -no  & -no- & -no  & thD  & SA0        & *        \cr
 365 &   -0.6 &   st & 0.43 & ex &  5.0 & ex & 0.46 &   7 & -no  & emDi & -no  & spH  & S          & *        \cr
 366 &    0.1 &   no & 0.75 & co &  3.0 & ex & 0.77 &   5 & -no  & -?Di & -no  & thD  & diE        & *        \cr
 367 &    0.2 &   no & 0.71 & ex & -0.2 & re & 0.80 &  11 & -no  & -no- & -no  & --  & unE        & *        \cr
 371 &        &      &      &    &      &    &      &     &      &      &      &      & S          & *        \cr 
 375 &   -- &   st & 0.39 & ex &  5.0 & ex & 0.51 &   3 & -no  & miDi & -no  & thD  & SA0        & *        \cr
 377 &   -0.5 &   st & 0.38 & ex &  4.0 & co & 0.47 &   3 & -no  & emDi & -no  & thD  & SA0        & *        \cr
 381 &   -0.1 &   st & 0.52 & ex &  1.9 & ex & 0.59 &  16 & -no  & emDi & -no  & thD  & SA0        &          \cr
 383 &        &      &      &    &      &    &      &     &      &      & spiP &      & S          & *        \cr
 384 &   -0.5 &   st & 0.54 & ex &  3.3 & co & 0.67 &   5 & --  &      & --  & spH  & unresolved & *        \cr
 385 &        &      &      &    &      &    &      &     &      &      &      &      & unresolved & *        \cr
 389 &   -0.8 &   st & 0.48 & ex &  7.0 & co & 0.57 &   3 & -no  & miDi & -no  & spH  & SA0        &          \cr
 393 &   -0.2 &   no & 0.71 & co & -1.0 & re &    - &   0 & -no  & -no- & -no  & --  & boE        & *        \cr
 397 &   -0.5 &   cl & 0.55 & ex &  3.4 & ex & 0.64 & 8:  & -no  & miDi & -no  & thD  & SA0        & *        \cr
 400 &   -0.6 &   no & 0.93 & co &  0.0 & co & 0.91 & 20: & -no  & -no- & -no  & --  & unE        & *        \cr
 408 &   -0.6 &   ft & 0.74 & co &  2.8 & co & 0.69 &  10 & -no  & emDi & -no  & thD  & diE/SA0    & *        \cr
 409 &   -- &   st & 0.21 & -- &  0.5 & -- & 0.45 &   3 & --   &      & -no  & -?-  & S          & *        \cr
 410 &   -0.6 &   st & 0.39 & ex &  8.1 & co & 0.61 &   4 & -no  & miDi & -no  & spH  & SA0        &          \cr
 413 &   -0.4 &   no & 0.85 & co &  1.7 & re & 0.85 &   0 & -no  & -no- & -no  & --  & diE        & *        \cr
 422 &   -0.3 &   st & 0.82 & re & -0.8 & re & 0.58 &  30 & bar? &      & -no  & --  & S          & *        \cr
 425 &   -0.6 &   st & 0.27 & co &  6.8 & co & 0.36 &   6 & -no  & emDi & -no  & thD  & SA0        &          \cr
 427 &   -0.2 &   no & 0.68 & co & -0.2 & re & 0.70 &   5 & -no  & -no- & -no  & --  & boE        &          \cr
 431 &   --&   st & 0.58 & ex &  3.6 & ex & 0.63 &  10 & -no  & emDi & -no  & spH  & SA0        & *        \cr
 432 &        &      &      &    &      &    &      &     &      &      &      &      & Irr        & *        \cr
 434 &   -0.5 &   st & 0.81 & re & -3.1 & ex & 0.81 &  21 & -no  &      & -no  & thD  & Irr        & *        \cr
 436 &   -0.4 &   st & 0.99 & ex &  0.0 & re & 0.91 &     &      & -- & -no  & -?-  & Pec      & *        \cr
 438 &   -0.5 &   st & 0.48 & ex &  2.0 & ex & 0.64 &  10 & -no  & miDi & -no  & spH  & SA0        &          \cr  
 439 &        &      &      &    &      &    &      &     &      &      &      &      & S          & *        \cr
 441 &   -0.4 &   st & 0.56 & re &  9.7 & ex & 0.31 &   5 & -no  & miDi & -no  & thD  & S          & *        \cr
}

\null\vfill\eject  
\null\vfill\eject 
\halign{\hfill#\quad&\hfill#\quad&\hfill#\quad&\hfill#\quad&\hfill#\quad&\hfill#\quad&\hfill#\quad&\hfill#\quad&
\hfill#\quad&\hfill#\quad&\hfill#\quad&\hfill#\quad&\hfill#\quad&#\hfill&\hfill#\quad\cr
 443 &   -0.7 &   cl & 0.69 & ex &  1.7 & ex & 0.83 &  10 & -no  & emDi & -no  & spH  & SA0        &          \cr
 448 &   -0.8 &   no & 0.95 & co &  0.0 & co & 0.95 & --  & -no  & -no- & -no  & --  & unE        & *        \cr
 451 &        &      &      &    &      &    &      &     &      &      &      &      & Irr        & *        \cr
 454 &        &      &      &    &      &    &      &     &      &      & spiP &      & S          & *        \cr
 457 &   -0.7 &   cl & 0.91 & ex &  2.0 & ex & 0.92 & --   & -no  & -?Di & -no  & --  & SA0        & *        \cr
 458 &   -0.3 &   no & 0.98 & co &  0.0 & co & 0.96 & --  & -no  & -no- & -no  & --  & unE        & *        \cr
 462 &        &      &      &    &      &    &      &     & bar? &      &      &      & S          & *        \cr
 471 &        &      &      &    &      &    &      &     &      &      & spiP &      & S          & *        \cr
\noalign{\medskip\hrule\medskip}
}            

\null\vfill\eject
\null\vfill\eject

\psfig{figure=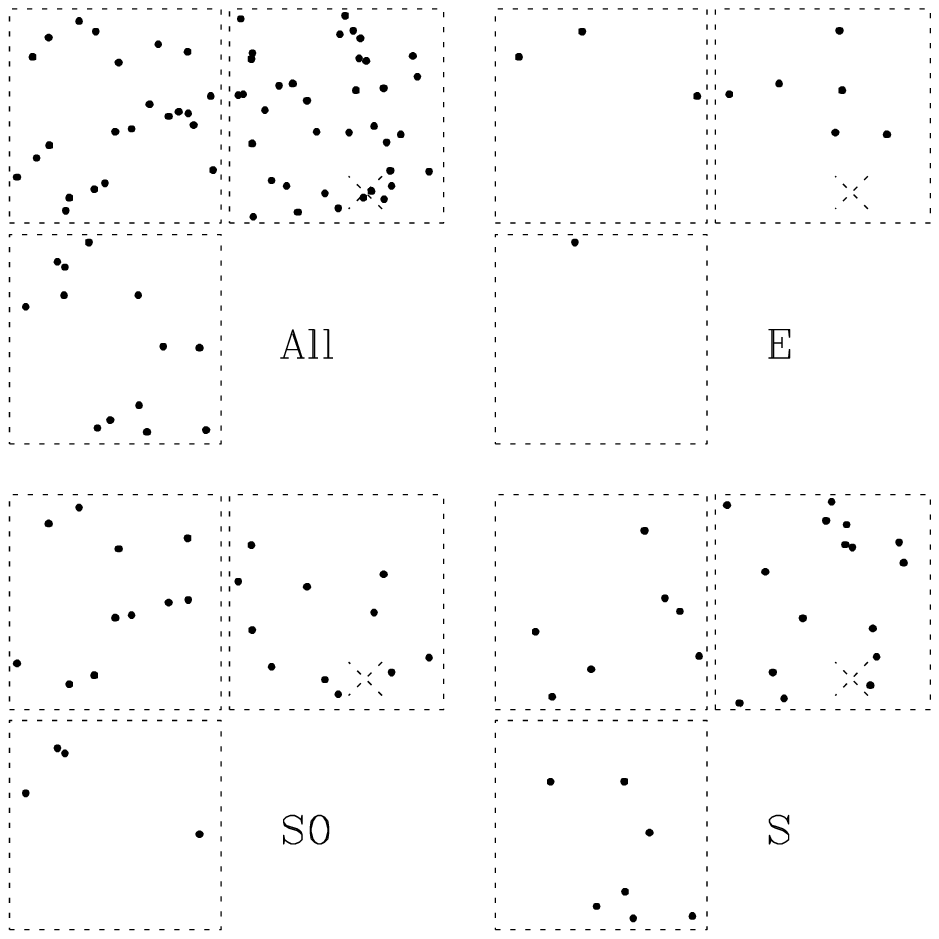,bbllx=25mm,bblly=130mm,bburx=175mm,bbury=230mm,width=21truecm}
\figure{1}{Spatial distribution of all galaxies of Cl0939+4713 and of each 
individual Hubble type. North is up and East is left. Dashed lines delimitate 
the field of view of the three CCDs of the Wide Field Camera 2 of the {\it 
HST}. The cross marks the adopted cluster center. The x,y coordinate system 
used in this work is rotated 20 degrees clockwise with respect to the 
celestial coordinate system, to be aligned with the 
preferred direction outlined by the distribution of Es.} 

\null\vfill\eject
\null\vfill\eject

\hbox{
\psfig{figure=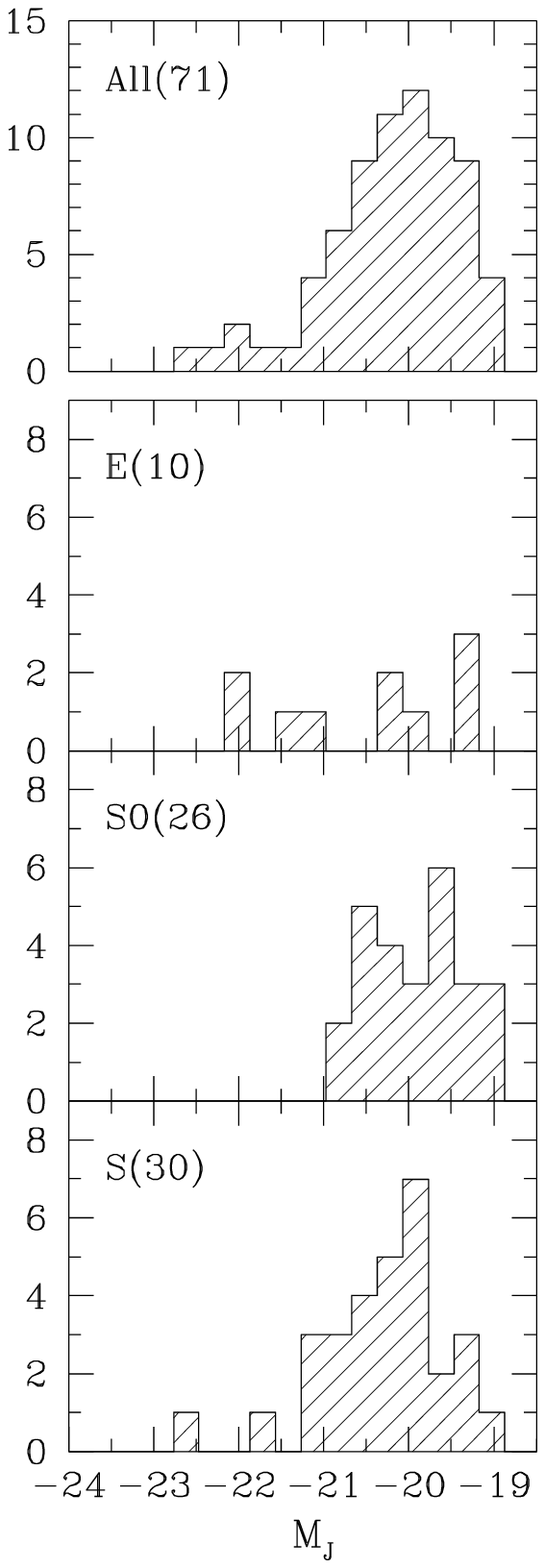,bbllx=45mm,bblly=75mm,bburx=105mm,bbury=240mm,width=6truecm,height=15truecm} \break
\psfig{figure=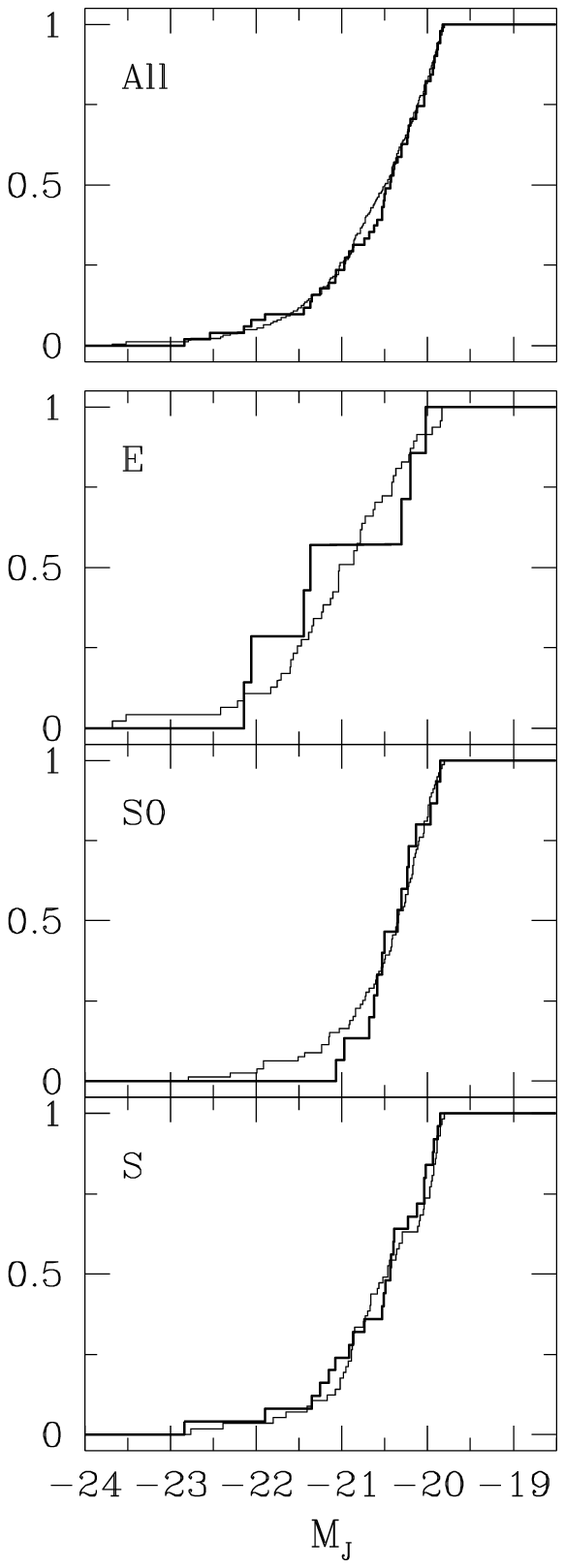,bbllx=45mm,bblly=75mm,bburx=105mm,bbury=240mm,width=6truecm,height=15truecm} \break
\psfig{figure=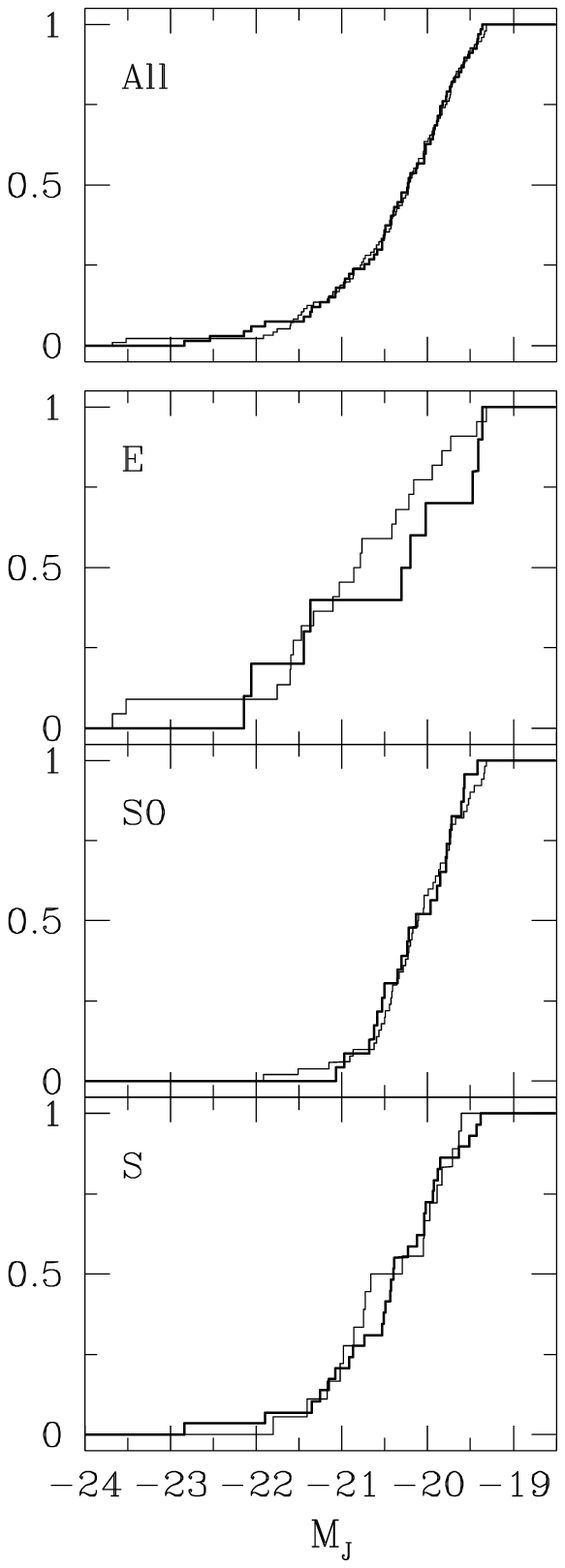,bbllx=45mm,bblly=75mm,bburx=105mm,bbury=240mm,width=6truecm,height=15truecm}}
\null\vfill
\figure{2a}{Left panel: $J$ luminosity function of all galaxies of Cl0939+4713 
and of each individual Hubble type. Central panel: comulative $J$ luminosity 
distribution for galaxies brighter than $J=-19.8$ for Cl0939+4713 galaxies 
(thick line) and Coma-whole (thin line). Right panel: same as the central 
panel, but for a limiting magnitude $J=-19.3$ mag and for the comparison 
sample Coma-center.} 

\null\vfill\eject
\null\vfill\eject

\hbox{
\psfig{figure=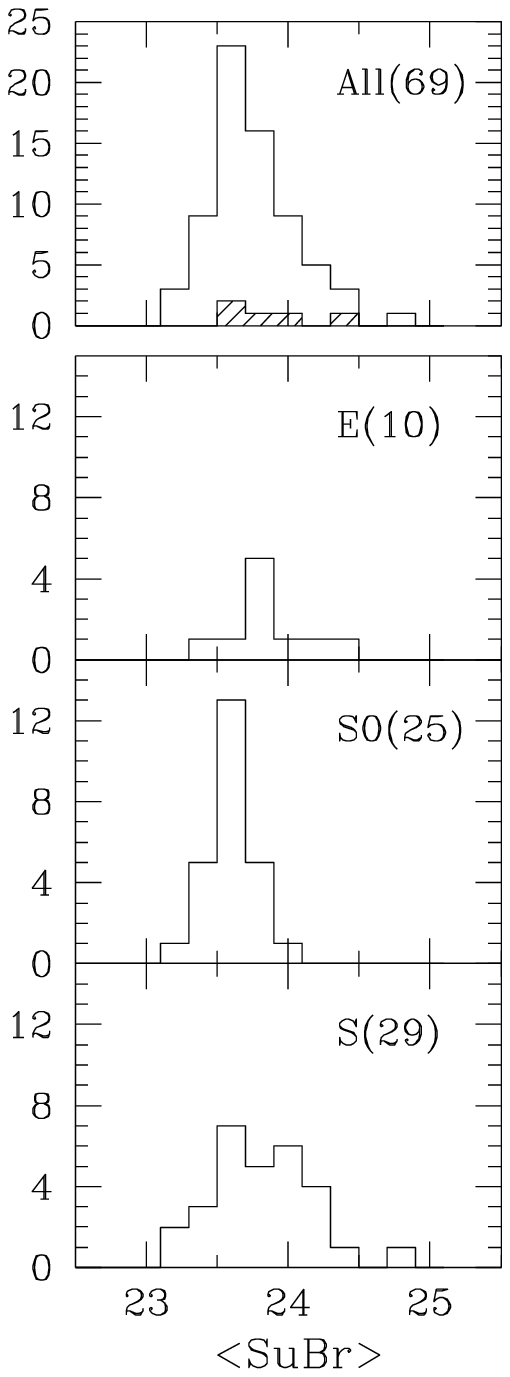,bbllx=50mm,bblly=105mm,bburx=105mm,bbury=240mm,width=6truecm,height=15truecm} \break
\psfig{figure=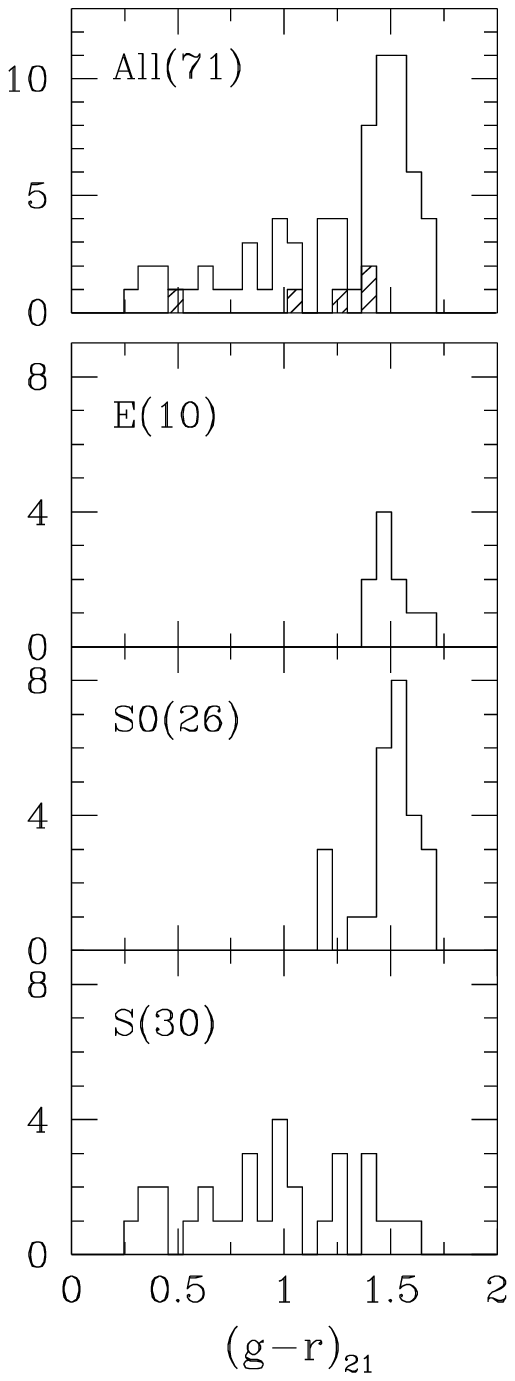,bbllx=50mm,bblly=105mm,bburx=105mm,bbury=240mm,width=6truecm,height=15truecm} \break
\psfig{figure=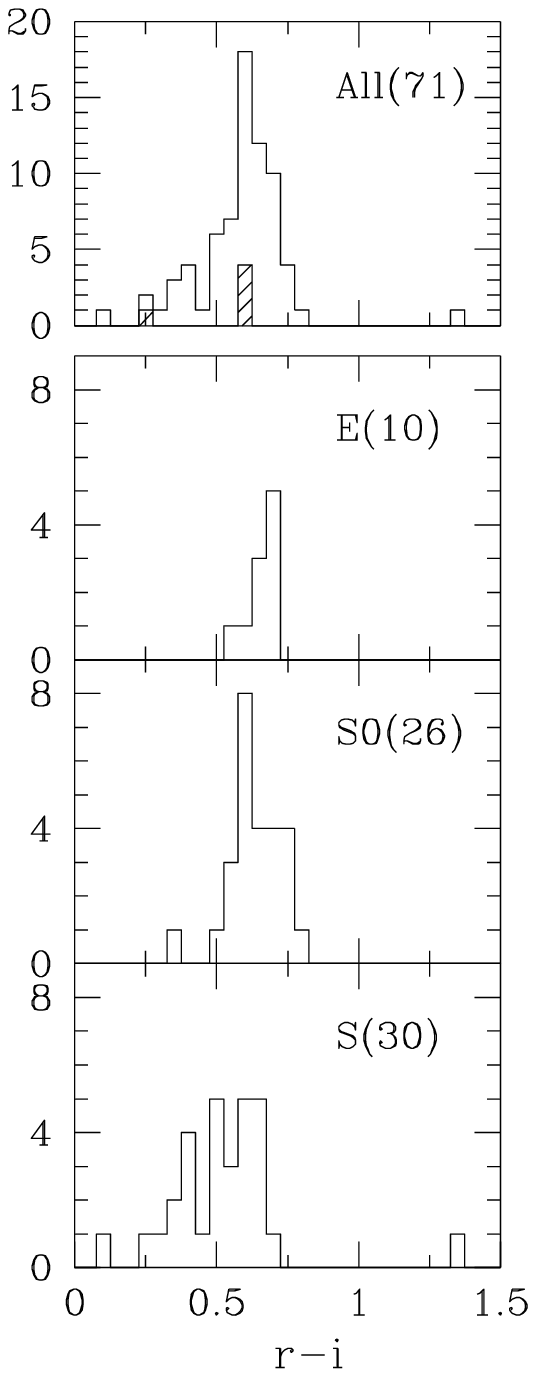,bbllx=125mm,bblly=105mm,bburx=180mm,bbury=240mm,width=6truecm,height=15truecm}}
\null\vfill
\figure{2b}{Mean apparent $J$ surface brightness, $(g-r)_{21}$ and $r-i$ color 
distributions of all galaxies of Cl0939+4713 and of each individual Hubble 
type.  The surface brightness is measured in mag arcsec$^{-2}$. $(g-r)_{21}$ is 
the $g-r$ color reduced to the magnitude $F702W=21$ taking into account the 
color-magnitude relation. The hatched histogram in the top panels shows the 
distributions for peculiar or unresolved galaxies.} 

\null\vfill\eject
\null\vfill\eject

\bye